\documentclass[aps,pra,showpacs,twocolumn,nofootinbib]{revtex4}

\usepackage{pstricks}

\def\bmath#1{\mbox{\boldmath$#1$}}

\begin{document}

\newcommand{\hgate}[6]{\rput[bl](#1,#2){\psframebox[framesep=0]{\parbox[t][#4][#6]{#3}{\begin{center}#5\end{center}}}}}

\title{Quantum Computational Complexity in the Presence of Closed
Timelike Curves}
\author{Dave Bacon}
\email[]{dabacon@cs.caltech.edu}
\affiliation{Institute for Quantum Information, California Institute
of Technology, Pasadena CA 91125, USA}
\affiliation{Department of Physics, California Institute of Technology, Pasadena CA 91125, USA}
                                                                                      \date{September 29, 2003}

\begin{abstract}
Quantum computation with quantum data that can traverse closed timelike curves represents a new physical model
of computation.  We argue that a model of quantum computation in the presence of closed timelike curves can be
formulated which represents a valid quantification of resources given the ability to construct compact regions
of closed timelike curves.  The notion of self-consistent evolution for quantum computers whose components
follow closed timelike curves, as pointed out by Deutsch [Phys. Rev. D {\bf 44}, 3197 (1991)], implies that the
evolution of the chronology respecting components which interact with the closed timelike curve components is
nonlinear. We demonstrate that this nonlinearity can be used to efficiently solve computational problems which
are generally thought to be intractable.  In particular we demonstrate that a quantum computer which has access
to closed timelike curve qubits can solve NP-complete problems with only a polynomial number of quantum gates.
\end{abstract}

\maketitle

The idea that {\em information is physical} has given rise to a series of discoveries which indicate that
physics has much to say about the foundations of computer science.  Computers which exploit coherent quantum
evolution remarkably offer computational speedups over computers which evolve
classically\cite{Shor:94a,Grover:96a}.  This discovery has lead to the development of a robust theory of
computation with quantum elements: the theory of quantum computation\cite{Nielsen:00a}.  Current theoretical
work\cite{Aharonov:97a,Gottesman:98a,Kitaev:97b,Knill:98a,Preskill:98a,Shor:96a} indicates that there is no
fundamental physical obstacle toward the construction of a working quantum computer.  The laws of physics
appear to allow quantum computation.

The realization that the physicality of information has a profound
effect on fundamental computer science challenges physics to
understand the computational power of different physical theories.
In this article we present an analysis of the consequences of one
such theory.  Morris, Thorne, and Yurtsever\cite{Morris:88a},
asked the question of whether the laws of physics allow for the
construction and maintenance of stable wormholes. The construction
of such wormholes would necessarily lead to spacetimes with closed
timelike curves(CTCs)\cite{Hawking:92a}.  Without a theory of
quantum gravity, however, there has been no conclusive resolution
of the question of whether nature allows for
CTCs\cite{Kim:91a,Hawking:92a,Deser:92a,Visser:96a}. Despite this
uncertainty, various authors have attempted to ascertain the
status of the initial value problem on spacetimes with
CTCs\cite{Friedman:90a,Deutsch:91a,Friedman:91a,Hartle:94a,Goldwirth:94a,Politzer:94a,Politzer:94b,Cassidy:95a,Hawking:95a}.
Of particular importance in this  initial value problem is the
notion of self-consistent
evolution\cite{Friedman:90a,Deutsch:91a}. Previous arguments
against the existence of CTCs which dictated that CTCs will always
lead to paradoxical evolution\cite{Hawking:73a} now appear to be
unfounded, especially in the context of quantum
theory\cite{Deutsch:91a,Hartle:94a}.  For a given specification of
initial data, there is always a self-consistent evolution of this
data which does not give rise to any of the typical ``patricidal
paradoxes'' usually associated with time travel.

In this article we examine the consequences of quantum computation in the presence of closed timelike curves.
This work is complementary to work done by Brun\cite{Brun:02a} who demonstrated that a model of classical
computation in the presence of CTCs could be used to solve hard computational problems in constant time.
However, the world is not classical, and the status of the classical initial value problem in the presence of
CTCs has no known generic solution\cite{Friedman:90a}.  Thus in Brun's model of computation, it is explicitly
possible to write down programs which have no self-consistent evolution.  Diverging from Brun's approach we
follow the formalism of Deutsch\cite{Deutsch:91a} who, soon after helping develop the theory of quantum
computation, applied this formalism to the question of computation in the presence of CTCs. Deutsch was able to
show that quantum computation in the presence of CTCs always allows self-consistent evolution.  The evolution
of chronology respecting systems interacting with systems which traverse CTCs, while locally unitary, Deutsch
showed, is globally nonlinear.  Nonlinearity in quantum computation has been shown by Abrams and
Lloyd\cite{Abrams:97a} to be a powerful tool for solving hard computational problems.  Both Deutsch and Brun
conjectured that quantum computation in the presence of CTCs could solve hard problems.  Here we show that this
is indeed correct by demonstrating that the nonlinearity allowed by quantum computation in the presence of CTCs
can be used to efficiently solve classically hard computational problems.  We present specific cases of quantum
evolution near CTCs which can be used to efficiently solve NP-complete problems.  The efficient solution of
such problems (the P=NP question) has long been doubted in classical computational complexity and it is also
believed that quantum computers alone do not efficiently solve these important computer science problems. If
nature allows for CTCs, then the theory of quantum computation in the presence of such CTCs provides for the
efficient solution of computational problems previously thought to be intractable and therefore represents one
of the most powerful physical models of computation known.

Given the extraordinary power of quantum computation in the presence of CTCs, however, one may wonder whether,
as is the case with the similarly powerful models of analog computation\cite{Vergis:86a}, this result is robust
in the presence of noise or whether noise destroys the effect we are exploiting to solve the hard
problem\footnote{This point was first brought to our attention by Patrick Hayden.}. This is particularly
worrisome because we use nonlinear evolution to achieve an exponential increase in the distinguishablity of two
nearly identical quantum systems: what is to keep the noise from growing exponentially along with this
distinguishablity?  We show, however, that the traditional methods of fault-tolerant quantum computation can be
used to overcome at least some of the problems raised in this context.  Thus, to the extent that the error
mechanisms we consider encompass realistic errors for the model of quantum computation in the presence of CTCs,
we find that a robust model of computation in the presence of CTCs can be formulated.

\section{Quantum complexity theory with closed timelike curves}  In physics, determination of the {\em
allowable} manipulations of a physical system is of central importance.  Computer science, on the other hand,
has arisen in order to {\em quantify} what resources are needed in order to perform a certain algorithmic task.
When one examines the computational consequences of a fundamental physical theory it is important that computer
science's quantification represent a reasonable application of physical resources.  One such quantification of
physical resources for a quantum computer is given by the quantum circuit model\cite{Deutsch:89a}.  In the
quantum circuit model, a series of gates are applied to a collection of qubits which have been prepared in an
input state and are then measured to obtain the computation's output.  In order to be a realistic model of
computation, it is usually assumed that there is some notion of locality among the quantum gates and further
that these gates are generated by few-qubit interactions.  The quantification of a quantum circuit model is
then classified by the manner in which the quantum gates are used.  There are various measures of complexity
within this model which can be used: one can use the total number of gates, the depth of the circuit, or the
breadth of the circuit.  That this is a good qualification of resources has been argued
elsewhere\cite{Deutsch:89a}.

In order to deal with CTCs within the quantum circuit model, we make
the simplifying assumptions enumerated by Deutsch\cite{Deutsch:91a}
and Politzer\cite{Politzer:94b}: (a) The region of CTCs is a compact
region of spacetime whose existence is generated by evolution from
initial conditions prior to this compact region. (b) Two types of
qubits in the quantum circuit model can be identified: those which
traverse CTCs and those which do not. (c) Unitary evolution between
the CTC qubits and the chronology respecting qubits is allowed. (d)
Measurement and preparation of the CTC qubits is not allowed (see
below however). (e) The evolution of the CTC qubits is determined by
self-consistency. Deutsch\cite{Deutsch:91a} enumerates reasons for
conjecturing that this model is universal in that any quantum
evolution in the presence of CTCs can be mapped onto this model.
Quantification of resources in this modified quantum circuit model
then follows the same lines of reasoning as in the unmodified
version.  Now, however, gates between all qubits (CTC qubits and
chronology respecting qubits) should be used in the quantification.
It should be noted that it this model, the CTC qubits are a resource
which cannot be reused: they are an expendable resource.  Further, it
is assumed that the resource cost of creating $n$ CTC qubits is not
exponential in $n$.  Finally, note that this quantification of
resources implies that the naive method of using a CTC to perform a
computation over and over again until the answer is arrived at is not
quantified as a tractable use of resources.  One could solve a hard
problem by trying out a solution to the problem, sending one's
computer back in time, attempting a different solution to the
problem, sending one's computer back and time, etc. until a solution
to the problem has been found.  While only a single wormhole could be
used for such an experiment and the total time required to obtain a
solution will be constant, the number of computers (i.e. spatial
resources) which need to exist to carry out this naive method is
exponential in the problem size (when one sends a computer back in
time, the previous versions of the computer still exist.) The goal of
this paper is to show that a better alternative to the naive approach
can actually be used to solve hard computational problems.

\section{Self-consistent Quantum Evolution}  
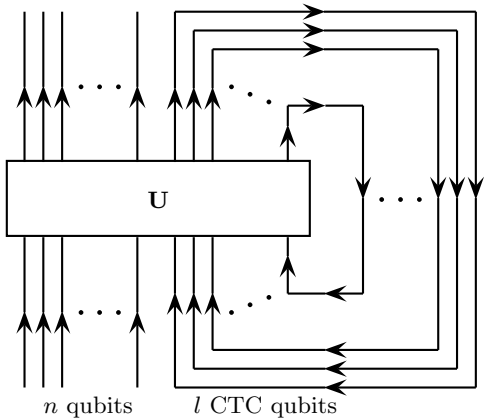
\begin{figure}[h]
\begin{pspicture}(0,0)(8,6)
 \hgate{1cm}{2.5cm}{4cm}{0.97cm}{${\bf U}$}{b}
 \psline[arrows=->,arrowscale=2](1.25,0.5)(1.25,1.52)
 \qline(1.25,1.5)(1.25,2.5)
 \psline[arrows=->,arrowscale=2](1.5,0.5)(1.5,1.52)
 \qline(1.5,1.5)(1.5,2.5)
 \psline[arrows=->,arrowscale=2](1.75,0.5)(1.75,1.52)
 \qline(1.75,1.5)(1.75,2.5)
 \psline[arrows=->,arrowscale=2](2.75,0.5)(2.75,1.52)
 \qline(2.75,1.5)(2.75,2.5)
 \psline[arrows=->,arrowscale=2](1.25,3.5)(1.25,4.52)
 \qline(1.25,4.5)(1.25,5.5)
 \psline[arrows=->,arrowscale=2](1.5,3.5)(1.5,4.52)
 \qline(1.5,4.5)(1.5,5.5)
 \psline[arrows=->,arrowscale=2](1.75,3.5)(1.75,4.52)
 \qline(1.75,4.5)(1.75,5.5)
 \psline[arrows=->,arrowscale=2](2.75,3.5)(2.75,4.52)
 \qline(2.75,4.5)(2.75,5.5)
 \psline[arrows=->,arrowscale=2](3.25,0.5)(3.25,1.77)
 \qline(3.25,1.75)(3.25,2.5)
 \qline(3.25,0.5)(5.25,0.5)
 \psline[arrows=<-,arrowscale=2](5.23,0.5)(7.25,0.5)
 \qline(7.25,0.5)(7.25,3.02)
 \psline[arrows=<-,arrowscale=2](7.25,3)(7.25,5.5)
 \qline(5.23,5.5)(7.25,5.5)
 \psline[arrows=->,arrowscale=2](3.25,5.5)(5.25,5.5)
 \qline(3.25,5.5)(3.25,4.48)
 \psline[arrows=->,arrowscale=2](3.25,3.5)(3.25,4.5)

 \psline[arrows=->,arrowscale=2](3.5,0.75)(3.5,1.77)
 \qline(3.5,1.75)(3.5,2.5)
 \qline(3.5,0.75)(5.25,0.75)
 \psline[arrows=<-,arrowscale=2](5.23,0.75)(7,0.75)
 \qline(7,0.75)(7,3.02)
 \psline[arrows=<-,arrowscale=2](7,3)(7,5.25)
 \qline(5.23,5.25)(7,5.25)
 \psline[arrows=->,arrowscale=2](3.5,5.25)(5.25,5.25)
 \qline(3.5,5.25)(3.5,4.48)
 \psline[arrows=->,arrowscale=2](3.5,3.5)(3.5,4.5)

 \psline[arrows=->,arrowscale=2](3.75,1)(3.75,1.77)
 \qline(3.75,1.75)(3.75,2.5)
 \qline(3.75,1)(5.25,1)
 \psline[arrows=<-,arrowscale=2](5.23,1)(6.75,1)
 \qline(6.75,1)(6.75,3.02)
 \psline[arrows=<-,arrowscale=2](6.75,3)(6.75,5)
 \qline(5.23,5)(6.75,5)
 \psline[arrows=->,arrowscale=2](3.75,5)(5.25,5)
 \qline(3.75,5)(3.75,4.48)
 \psline[arrows=->,arrowscale=2](3.75,3.5)(3.75,4.5)

 \psline[arrows=->,arrowscale=2](4.75,1.75)(4.75,2.27)
 \qline(4.75,2.25)(4.75,2.5)
 \qline(4.75,1.75)(5.25,1.75)
 \psline[arrows=<-,arrowscale=2](5.23,1.75)(5.75,1.75)
 \qline(5.75,1.75)(5.75,3.02)
 \psline[arrows=<-,arrowscale=2](5.75,3)(5.75,4.25)
 \qline(5.75,4.25)(5.25,4.25)
 \psline[arrows=<-,arrowscale=2](5.25,4.25)(4.75,4.25)
 \qline(4.75,4.25)(4.75,3.98)
 \psline[arrows=->,arrowscale=2](4.75,3.5)(4.75,4)

 \psdots[dotscale=0.5](2,1.5)(2.25,1.5)(2.5,1.5)
 \psdots[dotscale=0.5](2,4.5)(2.25,4.5)(2.5,4.5)
 \psdots[dotscale=0.5](4,1.5)(4.25,1.6)(4.5,1.7)
 \psdots[dotscale=0.5](4,4.5)(4.25,4.4)(4.5,4.3)
 \psdots[dotscale=0.5](6,3)(6.25,3)(6.5,3)

 \rput[bl](1.5,0.1){$n$ qubits}
 \rput[bl](3.5,0.1){$l$ CTC qubits}

\end{pspicture}
\caption{Pseudo spacetime diagram depicting quantum evolution in the presence of qubits which traverse closed
timelike curves.  The arrows indicate the forward direction of time for each individual qubit.  The $n$ qubits
on the left are the chronology respecting qubits and the $l$ qubits on the right are the qubits which traverse
closed timelike curves.} \label{fig:st}
\end{figure}

Consider a system of $n$ qubits, $n-l$ of which are qubits which evolve along
chronology respecting world lines and $l$ of which evolve along CTCs (see Fig.~\ref{fig:st}).  The Hilbert
space of these qubits is given by the tensor product ${\cal H} \equiv {\cal H}_A \otimes {\cal H}_B$ where
${\cal H}_A$ represents the chronology respecting qubits and ${\cal H}_B$ represents the qubits which traverse
CTCs. Input into the quantum circuit comes from the initial conditions of the chronology respecting qubits.  We
now make two assumptions whose validity we discuss below: (a) There is a temporal origin of the CTCs which can
be identified via the first interaction with the chronology respecting qubits.  (b) The initial state of the
chronology respecting and CTC qubits is initially uncorrelated.  Let ${\bf U}$ be the unitary evolution
operator of the entire system (made up of a series of gates), $\bmath{\rho}_{in}$ be the density matrix of the
chronology respecting qubits, and $\bmath{\rho}$ be the density matrix of the CTC qubits at the temporal origin
as defined above.  In order to avoid logical inconsistency of quantum theory, one must invoke the principle of
self-consistency: the state of the CTC qubits at the temporal origin should be the same as these same qubits
after the evolution ${\bf U}$. Mathematically, we have
\begin{equation}
\bmath{\rho}={\rm Tr}_A \left[ {\bf U} \left(\bmath{\rho}_{in} \otimes
\bmath{\rho} \right) {\bf U}^\dagger\right] \label{eq:consistent}
\end{equation}
where ${\rm Tr}_A$ represents the trace over ${\mathcal H}_A$. Deutsch\cite{Deutsch:91a} demonstrated that
there is always at least one solution $\bmath{\rho}$ to this self-consistency equation.  What to do with
multiple self-consistent solutions is discussed below.  Given a self-consistent evolution of the CTC qubits,
the output of the quantum circuit will be given by
\begin{equation}
\bmath{\rho}_{out}={\rm Tr}_B \left[ {\bf U} \left( \bmath{\rho}_{in}
\otimes \bmath{\rho} \right) {\bf U}^\dagger \right] \label{eq:output}
\end{equation}
where $\bmath{\rho}$ is a solution to Eq.~(\ref{eq:consistent}). Notice that the evolution from
$\bmath{\rho}_{in}$ to $\bmath{\rho}_{out}$ is possibly nonlinear due to the consistency condition: the
self-consistent solution to Eq.~(\ref{eq:consistent}) determines $\bmath{\rho}$ which in turn determines the
final mapping Eq.~(\ref{eq:output}).

 Returning now to our assumptions we first discuss a problem previously unaddressed in the
literature\cite{Deutsch:91a,Politzer:94a,Politzer:94b}.  Above we have assumed that there is a temporal origin
of the CTC evolution defined by the first interaction between the CTC qubits and the chronology respecting
qubits (who have a unique time ordering).  Now suppose that a gate ${\bf U}= \left( {\bf I} \otimes {\bf V}
\right) {\bf U}_0$ applied to this system.  The consistency condition is then
\begin{equation}
\bmath{\rho}={\rm Tr}_A \left[  \left( {\bf I} \otimes {\bf
V} \right) {\bf U}_0 \left(\bmath{\rho}_{in} \otimes
\bmath{\rho} \right) {\bf U}_0^\dagger\left( {\bf I} \otimes {\bf V}^\dagger \right)  \right]
\end{equation}
The temporal origin we have chosen for this consistency condition is now seen to be arbitrary in the following
sense. Express ${\bf V}$ as a product of two evolutions ${\bf V}={\bf V}_2 {\bf V}_1$.  The consistency
condition is
\begin{equation}
\bmath{\rho}_1={\rm Tr}_A \left[\left( {\bf I} \otimes {\bf
V}_2 {\bf V}_1 \right) {\bf U}_0 \left(\bmath{\rho}_{in} \otimes
\bmath{\rho}_1 \right) {\bf U}_0^\dagger \left( {\bf I} \otimes {\bf V}_1^\dagger {\bf
V}_2^\dagger \right)  \right]
\end{equation}
However there is no reason that the temporal origin should not be after ${\bf V}_1$ is applied such that the
consistency condition is really
\begin{eqnarray}
\bmath{\rho}_2&=&{\rm Tr}_A \left[ \left( {\bf I} \otimes {\bf V}_1
\right) {\bf U}_0 \left( {\bf I} \otimes {\bf V}_2 \right) \left(\bmath{\rho}_{in} \otimes
\bmath{\rho}_2 \right) \right. \nonumber \\
 &&\left . \left( {\bf I} \otimes {\bf V}_2^\dagger \right)
 {\bf U}_0^\dagger \left( {\bf I} \otimes {\bf V}_1^\dagger \right) \right] \
\end{eqnarray}
Via Deutsch's result, there are always self-consistent solutions to each of these different consistency
conditions. However, in general these self-consistent solutions may be different and even more disturbing is
that these different representations of the same physical process may lead to a different map between
$\bmath{\rho}_{in}$ and $\bmath{\rho}_{out}$.  This, however, is not the case.  The two self-consistent
solutions are related via a change of basis $\bmath{\rho}_2 = {\bf V}_2^\dagger \bmath{\rho}_1 {\bf V}_2$: the
superoperator on the non-CTC qubits is therefore the same superoperator.  Furthermore, the input-output
relationship is unaffected by a change of basis of the CTC qubits due to the cyclic nature of the trace in
Eq.(~\ref{eq:output}).  Therefore, while the choice of a temporal origin is arbitrary up to the temporal
ordering dictated by the chronology respecting system, every choice of a temporal origin results in the same
input output relationship for the chronology respecting qubits.

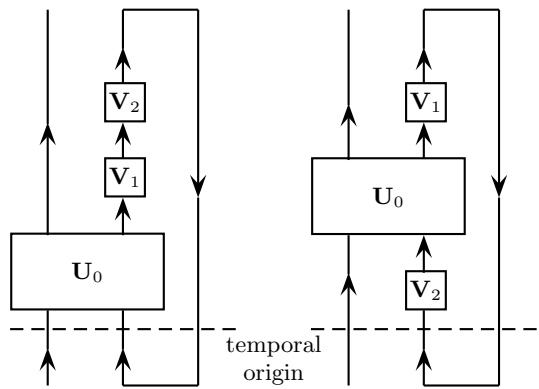
\begin{figure}[h]
\begin{pspicture}(0,0)(8,6)
 \hgate{4.5cm}{2.5cm}{2cm}{0.98cm}{${\bf U}_0$}{b}
 \hgate{5.75cm}{1.5cm}{0.5cm}{0.48cm}{${\bf V}_2$}{c}
  \hgate{5.75cm}{4cm}{0.5cm}{0.48cm}{${\bf V}_1$}{c}
  \psline[arrows=->,arrowscale=2](5,0.5)(5,2)
  \qline(5,1.98)(5,2.5)
  \psline[arrows=->,arrowscale=2](5,3.5)(5,4.25)
  \qline(5,4.23)(5,5.5)
  \psline[arrows=->,arrowscale=2](6,0.5)(6,1)
  \qline(6,0.98)(6,1.5)
    \psline[arrows=->,arrowscale=2](6,2)(6,2.5)
 \psline[arrows=->,arrowscale=2](6,3.52)(6,4)
 \psline[arrows=->,arrowscale=2](6,4.52)(6,5)
 \qline(6,4.98)(6,5.5)
 \qline(6,5.5)(7,5.5)
 \psline[arrows=->,arrowscale=2](7,5.5)(7,3)
 \qline(7,3)(7,0.5)
 \qline(7,0.5)(6,0.5)
 \psline[linestyle=dashed](4.5,1.25)(7.5,1.25)

 \hgate{0.5cm}{1.5cm}{2cm}{0.98cm}{${\bf U}_0$}{b}
 \hgate{1.75cm}{3cm}{0.5cm}{0.48cm}{${\bf V}_1$}{c}
  \hgate{1.75cm}{4cm}{0.5cm}{0.48cm}{${\bf V}_2$}{c}
  \psline[arrows=->,arrowscale=2](1,0.5)(1,1)
  \qline(1,0.98)(1,1.5)
  \psline[arrows=->,arrowscale=2](1,2.5)(1,4)
  \qline(1,3.98)(1,5.5)
  \psline[arrows=->,arrowscale=2](2,0.5)(2,1)
  \qline(2,0.98)(2,1.5)
    \psline[arrows=->,arrowscale=2](2,2.5)(2,3)
 \psline[arrows=->,arrowscale=2](2,3.52)(2,4)
 \psline[arrows=->,arrowscale=2](2,4.52)(2,5)
 \qline(2,4.98)(2,5.5)
 \qline(2,5.5)(3,5.5)
 \psline[arrows=->,arrowscale=2](3,5.5)(3,3)
 \qline(3,3)(3,0.5)
 \qline(3,0.5)(2,0.5)
 \psline[linestyle=dashed](0.5,1.25)(3.5,1.25)

\rput[tl](3.3,1.2){\begin{tabular}{c}temporal \\ origin \end{tabular}}

\end{pspicture}
\caption{The temporal origin problem.  While there is a notion of the time at which the CTC data first
interacts with the chronology respecting data, for evolution entirely acting on the CTC qubits before this
time, there are multiple ways to choose the temporal origin.  Here we show this ambiguity graphically with the
dashed line representing the time at which the consistency equation is applied.  As shown in the text, this
temporal origin ambiguity does not give rise to contradictions in the self consistency condition,
Eq.~\ref{eq:consistent}.}\label{fig:to}
\end{figure}

Next we turn to the assumption of an initially tensor product state $\bmath{\rho}_{in} \otimes \bmath{\rho}$.
Politzer \cite{Politzer:94b} has argued that this assumption is not the most general assumption.  The most
general initial state which produces the correct input state is one which satisfies $\bmath{\rho}_{in}={\rm
Tr}_B \left[ \bmath{\rho}_0 \right]$ where $\bmath{\rho}_0$ is the initial state of both the chronology
respecting and CTC systems. Politzer argues that it is wrong to assume that the state is initially a tensor
product because the CTC system has interacted with the chronology respecting system in the CTC system's past at
any given event along the CTC qubits history.  This withstanding, we note that there is always a factorizable
self-consistent solution\cite{Deutsch:91a}. Thus if non-factorizable solutions are also allowed, the difference
between these two must be in the initial value problem of the full chronology respecting plus CTC qubits. Thus
the model we consider, with factorizable inputs, is at least as powerful as the non-factorizable model of
Politzer.

\subsection{Example of the consistency requirement} Consider the evolution of two qubits, the first chronology
respecting and the second traversing a CTC under a controlled-phase gate followed by an exchange of the two
qubits: ${\bf U}=|00\rangle \langle 00| +|01\rangle \langle 10| + |10\rangle \langle 01| -|11\rangle \langle 11
|$ (we use a basis where $|ab\rangle$ is the chronology respecting qubit in state $|a\rangle$ and the CTC qubit
in state $|b\rangle$.)  The initial state of the system is given by the general input density operator
$\bmath{\rho}_{in}={1 \over 2} \left({\bf I}+ \vec{n} \cdot \vec{\bmath{\sigma}} \right)$ where $\vec{n}$ is
the Bloch vector $|\vec{n}| \leq 1$ and $\vec{\bmath{\sigma}}$ is the three vector of the Pauli operators
$\bmath{\sigma}_i$.  Similarly, a self-consistent CTC qubit state is $\bmath{\rho}={1 \over 2} \left( {\bf I}+
\vec{m} \cdot \vec{\bmath{\sigma}} \right)$.  The evolution of these qubits under ${\bf U}$ results in the
consistency conditions, Eq.~(\ref{eq:consistent}),
\begin{equation}
m_x=n_x n_z, \quad m_y=n_y n_z, \quad m_z=n_z.
\end{equation}
In this case we see that the density operator of the CTC qubit is
unique.  The output of the chronology respecting qubit can similarly
be calculated and found to be
\begin{equation}
\bmath{\rho}_{out}={1 \over 2} \left( {\bf I} + n_z^2 n_x
\bmath{\sigma}_x + n_z^2 n_y \bmath{\sigma}_y + n_z \bmath{\sigma}_z \right)
\end{equation}
Here we see that the evolution $\bmath{\rho}_{in} \rightarrow
\bmath{\rho}_{out}$ depends nonlinearly on the initial density matrix $\bmath{\rho}_{in}$.

\subsection{Multiple self-consistent evolutions}  In the previous example we have seen that there is a unique
self-consistent solution for the CTC qubit.  This, however, is not generally the case.  Consider the evolution
of the same two qubits, one chronology respecting and the other traversing a CTC, under a controlled-rotation
gate ${\bf U}=|00\rangle \langle 00| +|01\rangle \langle 01 | +|10\rangle \langle 10 | + i|11 \rangle \langle
11 |$. Again take the initial state of the chronology respecting qubit and the CTC qubit to be ${1 \over 2}
\left( {\bf I}+\vec{n} \cdot \vec{\bmath{\sigma}} \right) \otimes {1 \over 2} \left( {\bf I}+ \vec{m} \cdot
\vec{\bmath{\sigma}} \right)$.  In this case the consistency condition, Eq.~(\ref{eq:consistent}), yields the
condition
\begin{eqnarray}
m_x=0, \quad m_y=0, \quad m_z&=&{\rm unconstrained} \quad {\rm if}~n_z
\neq 1 \nonumber \\
m_x,m_y,m_z&=&{\rm unconstrained} \quad {\rm if }~n_z=1 \nonumber \\
\end{eqnarray}
Clearly the CTC qubit is unconstrained.  One may hope that while the
CTC qubits state is unconstrained, this does not affect the observable
evolution of the chronology respecting qubit.  However, the output
density matrix is given by
\begin{eqnarray}
\bmath{\rho}_{out}={1 \over 2} &&\left( {\bf I}+ \left(n_x {1 + m_z
\over 2} + n_y {-1+m_z \over 2} \right) \bmath{\sigma}_x \nonumber
\right. \\
&&\left.+ \left( n_x {1 - m_z \over 2}  + n_y {1 + m_z \over 2}\right) \bmath{\sigma}_y + n_z
\bmath{\sigma}_z \right) \nonumber \\
\end{eqnarray}
We therefore see that the output of this interaction is dependent on
the CTC qubit state.

Deutsch\cite{Deutsch:91a} has suggested that the self-consistent CTC density operator should be the chosen such
that this density operator maximizes the entropy $-{\rm Tr} \left[\bmath{\rho} \ln \bmath{\rho} \right]$.  We
should point out, however, that this solution to choosing which self-consistent solution may itself be
inconsistent: there may be multiple states maximizing the entropy which lead to different input output
evolutions of the chronology respecting qubits.  Further we note that there is another manner in which this
consistency paradox can be alleviated: one can assume that the freedom in the density matrix of the CTC systems
is an initial condition freedom.  One recalls that there are initial conditions which evolve into the CTC
qubit, i.e. the specification of conditions such that the compact region with CTCs is generated.  It is not
inconsistent to assume that some of the freedom in the initial conditions which produce this CTC qubit are
exactly the freedoms in the consistency condition.  Such a resolution to the multiple consistency problem puts
the impetus of explaining the ambiguity on an as yet codified theory of quantum gravity.  It is interesting to
turn this around and to ask if understanding the conditions for a resolution of the multiple consistency
problem can tell us something about the form of any possible theory of quantum gravity which admits CTCs?

Finally we note that not having a solution to the multiple self-consistent evolutions problem will not change
our result concerning quantum computational complexity in the presence of CTCs.  We will not encounter an
ambiguity of this form in our results: any solution to the multiple self-consistent evolutions problem is
compatible with our results.

\section{Efficient solutions to NP-complete problems using closed timelike curves}

Consider the following important example of a computation involving a single chronology respecting qubit and a
single CTC qubit.  In this example the unitary evolution of the two qubits is given by ${\bf U}=|00\rangle
\langle 00| + |10 \rangle \langle 01|+|11 \rangle \langle 10 | + |01 \rangle \langle 11|$ which corresponds to
the process of a controlled-NOT (controlled by the chronology respecting qubit) followed by swapping the two
qubits (this operation is also equivalent to two sequential controlled-NOT gates with alternating control
qubits.) Again assuming the initial state to be ${1 \over 2} \left( {\bf I}+\vec{n} \cdot \vec{\bmath{\sigma}}
\right) \otimes {1 \over 2} \left( {\bf I}+ \vec{m} \cdot \vec{\bmath{\sigma}} \right)$, the evolution of the
chronology respecting qubit is unambiguous if $n_x \neq 1$:
\begin{equation}
\bmath{\rho}_{out}={1 \over 2} \left( {\bf I} + n_z^2 \bmath{\sigma}_z \right)
\end{equation}
Examining this nonlinear evolution we can begin to see the power
afforded by the nonlinearity.  This map $n_z \rightarrow n_z^2$, when
repeated, can lead to an exponential separation of states on the Bloch
sphere which could normally not be distinguished.  Let ${\tt S}$
denote this map
\begin{equation}
{\tt S} \left[ {1 \over 2} \left( {\bf I}+ \vec{n} \cdot
\vec{\bmath{\sigma}} \right) \right] = \left \{ \begin{array}{ll} {1
\over 2} \left( {\bf I}+ n_z^2 \bmath{\sigma}_z \right) & {\rm
if}~n_x \neq 1 \\ {\rm ambiguous} &{\rm if}~n_x=1
\end{array} \right.
\end{equation}
which is not defined for $n_x=1$.

{\em Efficient solutions to NP-complete problems.--} The following
problem is NP-complete:
\begin{quotation}
{\bf Satisfaction (SAT): } Given a boolean function $f:\{0,1\}^n \rightarrow \{0,1\}$, specified in conjunctive
normal form, does there exist a satisfying assignment ($\exists x | f(x)=1$)?
\end{quotation}
In order to efficiently solve this problem we will make use of the
following oracle quantum gate acting on $n+1$ qubits
\begin{equation}
{\bf U}_f=\sum_{i=0}^{2^n-1} |i\rangle \langle i| \otimes \bmath{\sigma}_x^{f(i)}
\end{equation}
This gate can be constructed using only polynomial resources in the
size of the satisfaction problem.  Without quantifying exactly how
many resources are needed to enact this gate, we will simply show that
only a polynomial number of queries to this quantum gate can be used
in conjunction with ${\tt S}$ to solve the satisfaction problem.

The algorithm proceeds as follows.  First the state $|\psi_0\rangle=\left({1
\over \sqrt{2^n}} \sum_{i=0}^{2^n-1} |i\rangle \right) \otimes
|0\rangle$ is prepared and acted upon by ${\bf U}_f$.  This prepares
the state ${1 \over \sqrt{2^n} }\sum_{i=0}^{2^n-1} |i \rangle \otimes
|f(i)\rangle $.  The reduced density operator of the final qubit is
now given by
\begin{equation}
\bmath{\rho}_0={1 \over 2} \left( {\bf I} + \left(1 - {s
\over 2^{n-1}} \right)
\bmath{\sigma}_z \right) \label{eq:out}
\end{equation}
where $s$ is the number of satisfying solutions to $f(x)=1$.  We can assume that $s \neq 2^n$ for, if $s=2^n$,
then we could easily solve this case by randomly querying a value of $f(x)$.  We therefore wish to distinguish
between $s=0$ and $0<s<2^n$.

Let $\gamma$ denote the $\bmath{\sigma}_z$ component of $\bmath{\rho}_0$. Initially this component is $\gamma=1
- {s \over 2^{n-1}}$. After applying the gate ${\tt S}$ $p>1$ times, the component of the $\bmath{\sigma}_z$
evolves to $\gamma_p= \left( 1- {s \over 2^{n-1}} \right)^{2^p}$.  Notice if $s=0$, $\gamma_p=1$, and if
$0<s<2^n$ then $\gamma_p$ tends to $0$ exponentially fast in $p$.  After performing ${\tt S}$ $p$ times, one
measures the qubit in the $\bmath{\sigma}_z$ basis.  This whole procedure is then repeated $q$ times (for a
total of $pq$ queries to ${\bf U}_f$).  If any of the measurements during these $q$ runs yields
${\sigma}_z=-1$, then the algorithm outputs that there is a satisfying input.  If none of the measurements
yields $\sigma_z=-1$, then the algorithm outputs that there is no satisfying input.  When there is no
satisfying clause, this algorithm will always get the answer correct.  When there is a satisfying clause, the
algorithm will incorrectly identify this has having no satisfying clause with a probability
\begin{equation}
P_{fail}={1 \over 2^q} \left( 1+ \left( 1- {s \over 2^{n-1}}
\right)^{2^p} \right)^q
\end{equation}
With $p$ and $q$ polynomial in $n$ the probability of this algorithm failing is therefore exponentially small.

We therefore see that using the nonlinearity provided by the gate $\tt S$ we can amplify the probability in the
quantum computer such that NP-complete problems can be efficiently computed.  Via the definition of
NP-completeness we therefore have shown that any problem in the class NP can be efficiently solved by our
algorithm.

\section{Error Correction}  While we have demonstrated that an error-free quantum computer can, in the
presence of CTCs, solve a hard problem, we do not have a fully convincing argument unless we can argue that the
presence of noise or faulty components does not destroy this result.  Here we argue that that the presence of
noise in the system will not destroy our result.

Recall from the theory of fault tolerant quantum
computation\cite{Aharonov:97a,Gottesman:98a,Kitaev:97b,Knill:98a,Preskill:98a,Shor:96a} that a quantum circuit
containing $p(n)$ gates can be simulated with a probability of error $\epsilon$ using
$n^\prime(p(n),\epsilon)=O\left({\rm poly} \left(\log \left({p(n)\over \epsilon} \right)\right) p(n) \right)$
gates which fail with probability $p<p_{threshold}$ for some fixed $p_{threshold}$.   One way to interpret this
result is to say that if we want to define our density matrix up to probabilities of outcomes given by
$\epsilon$, then we require an error correcting overhead $n^\prime(p(n),\epsilon)$ gates.  We would like to use
fault-tolerant methods on our construction of $\tt S$. The simulation of a quantum circuit in fault tolerant
constructions occurs via encoding the quantum information into appropriate error correct code states and by
acting with particular operations which fault-tolerantly act on this encoded quantum information. For $\tt S$
constructed above, this means that we need to have both the chronology respecting and the CTC systems evolve
with encoded quantum information and for fault tolerant gates to act on both of these systems. For the
chronology respecting qubits we can clearly arrange for the appropriate encoding.  For the CTC qubits, however,
it is not clear how to arrange for the appropriate encoding to occur.  How can we use fault tolerant encoded
methods when we cannot reach in to the CTC qubits and perform the appropriate encoding?

He we sketch a method to overcome this encoding difficulty.  To simplify our discussion we will focus on fault
tolerant methods which use the class of stabilizer codes known as Calderbank, Shor, and Steane (CSS)
codes\cite{Calderbank:96a,Steane:96a}. These codes are particularly nice for our construction of $\tt S$
because the encoded controlled-not operation can be implemented transversely (and hence fault-tolerantly) by a
series of controlled-not's from the encoded control qubits to the encoded target bits. The first observation
which will allow us to perform fault tolerant methods on the chronology respecting and CTC qubits is to note
that the consistency and evolution equations, Eq.~\ref{eq:consistent} and Eq.~\ref{eq:output}, when considered
over the error correcting codespace will yield identical evolution and consistency for the encoded quantum
information as for the identical unencoded evolution when the fault tolerant operations preserve the
codespaces.  The transversal controlled-not for the CSS codes preserve the error correcting codespaces.  Thus
if in the unencoded evolution the CTC qubits are forced by consistency to be in the state $\bmath{\rho}$, then
for the same encoded evolution, there is a self-consistent solution over the error correcting subspace which
corresponds to the encoded version of $\bmath{\rho}$.  The main problem then is that there may be other
self-consistent evolutions which involve CTC qubits in states outside of the error correcting codespace.

The second observation we need is that there is a degeneracy in stabilizer coding which allows us to consider
the full Hilbert space of the CTC qubits as divided into different {\em equivalent} error correcting code
spaces.  In particular, the error correcting subspace normally used corresponds to considering the subspace
spanned by $+1$ eigenvalue eigenstates of a set of operators known as the generators of the stabilizer
group\cite{Gottesman:97a}.  However, one could equally well work with the subspace spanned by any fixed $\pm1$
eigenvalue eigenstates of the generators of the stabilizer group.  Knowing the $\pm 1$ eigenvalues defines a
codespace which is of equivalent error correcting capacity as the all $+1$ eigenvalue codespace.  The
operations which we perform for the $\pm 1$ eigenvalue codespace will be different than those if we used the
all $+1$ eigenvalue codespace, but there is always an equivalent set of operations for this other $\pm 1$
eigenvalue codespace.  Thus if we could encode into any one of these $\pm1$ eigenvalue codespaces, then, via
our first observation, we could again guarantee correct encoded evolution.  Finally we need the fact that all
of the $\pm 1$ codespaces together span the entire space of unencoded qubits.

The three observations above allow us to perform the following procedure on our CTC qubits which effectively
allows us to avoid the encoding problem on the CTC qubits.  First, we perform fault-tolerant measurements of
the stabilizer generators on the CTC qubits which put the result of these measurements in these chronology
respecting qubits (as is done fault-tolerantly in \cite{Gottesman:97a}.) Then, for all further operations, we
classically control on this encoded measurement result the appropriate evolution (this needs to be done with a
fault tolerant construction.)  Thus, the self-consistent evolution will always be the appropriate encoded
self-consistent evolution and the evolution will be the proper encoded evolution, but for the CTC qubits over a
particular $\pm 1$ eigenvalue codespace.

Return now to the issue of using fault tolerance for quantum computation in the presence of CTC qubits.  Notice
that in our algorithm for efficiently solving NP-complete problems, we need to distinguish the $s=0$ state
$\bmath{\rho}(s=0)={1 \over 2} \left( {\bf I}+\bmath{\sigma}_z\right)$ and the $s=1$ state $\bmath{\rho}(s=1)={
1 \over 2} \left( {\bf I} + \left( 1-{1 \over 2^{n-1}} \right) \bmath{\sigma}_z \right)$. Since the trace
distance\cite{Fuchs:96a} between these two states is given by $D(\bmath{\rho}(s=0),\bmath{\rho}(s=1))={1 \over
2} {\rm Tr} \left( | \bmath{\rho}(s=0) - \bmath{\rho}(s=1)| \right)= {1 \over 2^{n-1}}$ then we clearly need to
be able to use error correction to maintain at least the probability difference $\epsilon={1 \over 2^n}$. Using
standard error correction this can be done using $O(\log\left(p(n) 2^n \right) p(n))=O(\log\left(p(n) n) p(n)
\right)$ faulty gates (operating below the threshold.)  This is simply a polynomial increase in the size of the
quantum circuit and therefore does not significantly slow down our CTC algorithm for NP problems.

A slightly more worrisome type of error is as follows.  Suppose that in our algorithm the $\bmath{\rho}(s=0)$
state has a component of $\bmath{\sigma}_z$ which is different than $+1$, $\tilde{\bmath{\rho}}(s=0)={1 \over
2} \left( {\bf I}+(1-\mu) \bmath{\sigma}_z \right)$, due to some physical noise process. Then if we apply ${\tt
S}$ $p$ times to this state, it is possible, for large enough $\mu$ that our algorithm will incorrectly
identify that the function has a satisfying assignment when, in fact the function does not.  Suppose we run the
first phase of our algorithm $p=n$ times.  The state $\tilde{\bmath{\rho}}$ will then have a $\bmath{\sigma}_z$
component of $\left(1-\mu\right)^{2^n}$.  Suppose $\mu > {b \over 2^{n^c}}$ for some constants $b$ and $c$ for
large $n$. Then for large $n$,
\begin{equation}
\left(1-\mu\right)^{2^n} \geq e^{-2^n \over 2^{n^c} +1 } \geq 1-{2^n \over {2^n}^c+1}.
\end{equation}
If we choose some fixed $c > 1$, then for large $n$ the $\bmath{\sigma}_z$ component is exponentially close to
$1$.  Therefore repeated applications of ${\mathcal S}$ will improperly identify the $s=0$ case with only an
exponentially small probability.  This implies that we need only protect our system to accuracy
$\epsilon=O\left({1 \over 2^{n^c}} \right)$ for some fixed $c$.  This can be done using the threshold theorem
using $O(\log(p(n))n^cp(n))$ gates, representing a polynomial slowdown, and thus error correction can be used
to correct his form of error.  While it is true that the nonlinearity we use to solve hard problems
exponentially separates quantum states, it appears that we can design quantum error correcting circuits whose
noise does not suffer a similar blowup.  We have considered only limited errors in this paper, and then only as
a sketch as to how more general errors can be dealt with in the presence of nonlinear quantum gates.  It
remains a challenge bring a fully rigorous treatment of errors in our model to be certain that our model is
robust in the presence of noise.  We have shown, however, that the problems for which using nonlinearly at
first sight appear to be problematic are not problematic and we are thus hopeful that such a full rigorous
theory could be developed.

\section{Conclusion}

Since computation is physical, we need to examine physics in order to form the foundations of a theory of
computational complexity.  There are two direct ways in which one can go beyond the standard model of quantum
computation for which physics might have something new to say about computational complexity.  One possible
manner to go beyond the standard model would occur if quantum theory needs to be replaced by a more fundamental
theory of the evolution of physical systems.  For example, proposals for deterministic nonlocal
theories\cite{tHooft:01a,Smolin:02a}, might offer different computational complexities if the fundamental
distinctions which make these models different from quantum theory are accessible.  Another example is provided
by Hawking's conjecture\cite{Hawking:76a,Preskill:92a} that quantum theory must be modified to solve the
information paradox in black hole thermodynamics.  The other path beyond the standard model of quantum
computation is if the physical theories which are laid on top of quantum theory possess a computational power
differing from the current understanding of these theories.  For example, it is not entirely clear whether or
not current versions of quantum gravity\cite{Smolin:03a} provide physics which is computationally equivalent to
the standard model. If the physical theory of gravity is itself to provide a picture of spacetime, how does
this modify the theory of computational complexity which is grossly constrained by the geometry of the
computer?  In this paper we have consider a possible hybrid method which will produce computational complexity
which appears to be stronger than the standard model of quantum computation. We have considered that the
possible existence of closed timelike curves might follow from a quantum theory of gravity and using the
structure uncovered by Deutsch for how quantum theory itself must be modified in the presence of such curves to
solve hard computational problems. There are of course many open issues left to be addressed, not the least
whether a theory of quantum gravity exists which is compatible with CTCs.  However, there are interesting
possibilities which might also warrant consideration solely for their theoretical usefulness.  For example, in
computer science, the difference between space and time complexity is poorly
understood\cite{Papadimitriou:95a}. Complexity classes which quantify reasonable amounts of spatial resources
appear to be more powerful than complexity classes which quantify amounts of temporal resources.  An obvious
reason for this difference lies in the fact that spatial resources may be reused while temporal resources only
be used once.  Clearly if nature allows CTC's this obstruction is partially remove. It is interesting to
speculate that computational complexity with CTCs can lead to a simplified theory of computational complexity.

Finally, we would not be honest if we did not end this paper with the caveat that this work is at best a
creature of eager speculation. Without a theory of quantum gravity, we cannot know whether CTCs can exist let
alone whether they can be generated within the confines of the such a theory.  Practical considerations are
humorous at best. The surprising answer that quantum computation in the presence of CTCs is a powerful new
model of quantum computation gives us reason, however, to pause and ponder the implications.

\begin{acknowledgements}
We thank Patrick Hayden for bringing to our attention concerns about
error correction when using nonlinear quantum evolution.  We also
acknowledge useful conversations and communications with Todd Brun,
Mike Cai, Aaron Denney, David Deutsch, Julia Kempe, John Preskill and
Benjamin Toner. This work was supported in part by the National
Science Foundation under grant EIA-0086038 through the Institute for
Quantum Information.
\end{acknowledgements}

\bibliography{bigref}

\end{document}